\documentclass[aps,pra,superscriptaddress,
longbibliography,preprint]{revtex4-1}
\usepackage[utf8]{inputenc}
\usepackage{graphicx}
\usepackage{graphics}
\usepackage{dsfont}
\usepackage{bm}
\usepackage{amsmath}
\usepackage{amsfonts}
\usepackage{mathrsfs}
\usepackage{braket}
\usepackage[capitalize]{cleveref}
\usepackage{units}
\usepackage{color}

\begin{document}
\title{On the energy conversion efficiency of the bulk photovoltaic effect}
\author{Andreas Pusch}
\email{andreas.pusch@gmx.net}
\affiliation{School of Photovoltaic \& Renewable Engineering, UNSW Sydney, Kensington 2052, Australia}
\author{Udo R{\"o}mer}
\affiliation{School of Photovoltaic \& Renewable Engineering, UNSW Sydney, Kensington 2052, Australia}
\author{Dimitrie Culcer}
\affiliation{School of Physics, UNSW Sydney, Kensington 2052, Australia}
\author{Nicholas J. Ekins-Daukes}
\affiliation{School of Photovoltaic \& Renewable Engineering, UNSW Sydney, Kensington 2052, Australia}

\begin{abstract}
The bulk photovoltaic effect (BPVE) leads to directed photo-currents and photo-voltages in bulk materials.
Unlike photo-voltages in p-n junction solar cells that are limited by carrier recombination to values below the bandgap energy of the absorbing material, the BPVE photo-voltages have been shown to greatly exceed the bandgap energy.
Therefore the BPVE is not subject to the Shockley-Queisser limit for sunlight to electricity conversion in single junction solar cells and experimental claims of efficiencies beyond this limit have been made.
Here, we show that BPVE energy conversion efficiencies are, in practice, orders of magnitude below the Shockley-Queisser limit of single junction solar cells and are subject to different, more stringent limits.
The name BPVE stands for two different fundamental effects, the shift current and the injection current.
In both of these, the voltage bias necessary to produce electrical energy, accelerates both, intrinsic and photo-generated, carriers.
We discuss how energy conservation alone fundamentally limits the BPVE to a bandgap-dependent value that exceeds the Shockley Queisser limit only for very small bandgaps.
Yet, small bandgap materials have a large number of intrinsic carriers, leading to high conductivity which suppresses the photo-voltage.
We discuss further how slightly more stringent fundamental limits for injection (ballistic) currents may be derived from the trade-off between high resistivity, needed for a high voltage, and long ballistic transport length, needed for a high current.
We also explain how erroneous experimental and theoretical claims of high efficiency have arisen.
Finally, we calculate the energy conversion efficiency for an example 2D material that has been suggested as candidate material for high efficiency BPVE based solar cells and show that the efficiency is very similar to the efficiency of known 3D materials.
\end{abstract}

\maketitle

\section{Introduction}

One can generally distinguish between two different effects that can give rise to photo-voltages upon illumination, the barrier layer photovoltaic effect and the bulk photovoltaic effect (BPVE) \cite{Tauc1957}. In the barrier layer photovoltaic effect, the voltage is a consequence of a quasi-Fermi level separation between electrons in the conduction and holes in the valence band that develops upon illumination. A current flows due to a broken spatial symmetry in conductivity of electrons and holes \cite{WurfelBook}, often achieved by p-n junctions.
This barrier layer effect forms the basis of all current commercial photovoltaic solar energy harvesting devices.

The quasi-Fermi level separation increases the product of the carrier concentrations in conduction and valence band leading to enhanced radiative recombination \cite{Wurfel1982}.
This unavoidable recombination limits the open circuit voltage to a value below the optical bandgap of the absorber material leading to the Shockley Queisser efficiency limit \cite{Shockley1961} for single junction solar cells.
Existing technology has already very closely approached this single junction limit of $33.7\%$ for AM1.5G illumination \cite{Green2022}.
Further major improvements, reaching far above $40\%$ energy conversion efficiency, are being made by elaborate layering of several junctions of materials with decreasing bandgap into a series-connected device \cite{Geisz2020}.

The photo-voltage in the BPVE has a fundamentally different origin than the voltage in the barrier layer effect.
It is either due to an instantaneous shift in the charge distribution upon absorption of light, the so-called shift current \cite{vonBaltz1981} stemming from the interband Berry connections \cite{Young2012} or due to an injection of a carrier distribution that is asymmetric in momentum space and gets transported ballistically \cite{Zhang2019}.
This injection occurs either directly through the photoexcitation process or involves scattering with impurities or phonons.
In practice, several contributions are often present simultaneously \cite{Sturman2020}.
As a consequence of this microscopic current, a voltage builds up in the resistive material that is inversely proportional to the conductivity and proportional to the photo-current.
In the shift current as well as the injection/ballistic current case, the BPVE can be described as a second order nonlinear effect \cite{Sipe2000} quantified by the nonlinear conductivity tensor $\sigma_{ijk}$ for the interaction of positive and negative frequencies to generate a zero-frequency current density
\begin{equation}
    \label{eq:SecondOrderShift}
    J^{i} = \sigma_{ijk}(0;\omega,-\omega) E_{j} E_{k} \, . 
\end{equation}
The second order nonlinear response vanishes for centro-symmetric materials, so that the BPVE requires the breaking of inversion symmetry.

It has been established that the photo-voltage that can be reached with the BPVE is higher than the band gap of the material \cite{Yang2010,Yang2017}.
This is clear evidence that the Shockley-Queisser limit for solar energy conversion based on the barrier layer effect does not apply to the BPVE.
Many researchers have taken this to mean that the BPVE can lead to more efficient photovoltaic devices \cite{Tan2016,Cook2017}.
A controversial experimental claim of high efficiency has been made \cite{Spanier2016} which has received criticism for its normalization procedure \cite{Kirk2017}.
However, a calculation of an alternative energy conversion limit, whether for monochromatic or broadband illumination, is lacking.

Here we derive a simple formula for the energy conversion efficiency of the BPVE that is independent of device structure.
Using the example of ballistic currents, we explain how energy conversion efficiency limits may arise from this formula.
We then investigate recent claims of high energy conversion efficiency and explain why they are flawed.
Finally, we estimate the conversion efficiency for a monolayer of GeSe and show that caution has to be used when interpreting average 3D values of 2D susceptibilities.

\section{The conversion efficiency from material parameters}

\begin{figure}
    \centering
    \includegraphics{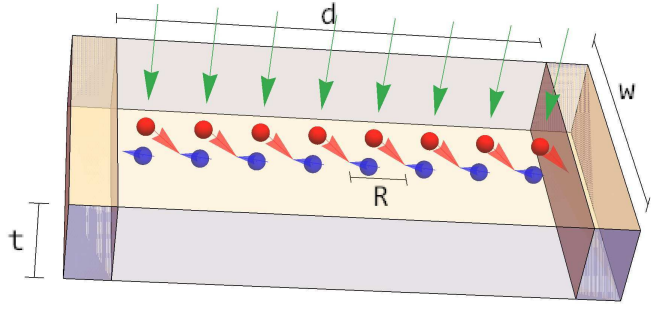}
    \caption{An illustration of the current in a device made of a bulk semiconductor and two electrodes at the sides of the device. The average shift of the charges (electrons and holes) per absorption event of the incoming light (green arrows) is denoted as $R$. This shift may occur due to ballistic transport or due to the shift vector.}
    \label{fig:BPVEconcept}
\end{figure}

The energy conversion efficiency of a photovoltaic device is conventionally defined by the ratio between the electrical power, i.e. the product of the current $I$ and the voltage $V$ at the maximum power point of the current-voltage characteristic (J-V curve), and the total power $\mathcal{I} A_{i}$ contained in the incoming light field, i.e.,
\begin{equation}
    \label{eq:efficiencyFundamental}
    \eta_{PV} = \frac{I V}{\mathcal{I} A_{i}} \, ,
\end{equation}
with the light intensity $\mathcal{I}$ and the illuminated area $A_{i}$.
Maximum power and efficiency are found by maximising the $I$ $V$ product along a I-V curve. 
The BPVE can have different device geometries according to the direction of the photo-generated current, which may depend on the alignment of the material with the incoming light field.
Here, we assume a device in which the electrical contacts are placed on either side of an illuminated area and the relevant current density flows orthogonal to the Poynting vector of the incoming light (see Figure \ref{fig:BPVEconcept}).
A geometry in which the Poynting vector and the current density are parallel to each other is also possible but an important, but sometimes neglected, subtlety of the efficiency calculation is lost in that geometry.
Namely, in the orthogonal geometry of Figure~\ref{fig:BPVEconcept} the illuminated area $A_{i}=d*w$ with the distance between contacts $d$ and the width of the contacts $w$ is not the same as the area through which the current flows, given by $A_{c} = w*t$, with the thickness $t$ of the active material.

The total steady state short circuit current for polarized monochromatic light of frequency $\omega$ can be written as a function of the average charge displacement upon absorption of a photon, denoted as $R$, which corresponds to the average over the shift vector in the shift current case and the directed average over the ballistic transport distance in the injection current case
\begin{equation}
    \label{eq:JscFromR}
    I_{sc} = q A_{i} \frac{R}{d} a \phi \, .
\end{equation}
Here, the product of absorptivity $a$ (with values of $0 \leq a < 1$), photon flux $\phi$ and illumination area $A_{i}$ describe the number of photons absorbed per unit time. On average it requires $d/R$ absorption events to move one unit of charge $q$ from one contact to the other.
This is in sharp contrast to barrier layer photovoltaics in which one absorbed photon ideally leads to one electron-hole pair reaching the contacts.
To map the short circuit current to tabulated material parameters \cite{Sturman1992}, we can rewrite the expression for the short circuit current in terms of the Glass coefficient $G$, with $G = I_{sc}/(\mathcal{I} w)$, given by
\begin{equation}
    G = \frac{q R}{\hbar \omega} \, ,
\end{equation}
resulting in
\begin{equation}
    \label{eq:JscFromG}
    I_{sc} =  A_{i} \frac{G \hbar \omega}{d} a \phi =  A_{i} \frac{G}{d} a \mathcal{I} \, .
\end{equation}
Note that both $G$ and $R$ are tensorial quantities, depending on polarization direction and direction of current measurement, but for the purpose of deriving limits we are only interested in their largest components.

\begin{figure}
    \centering
    \includegraphics{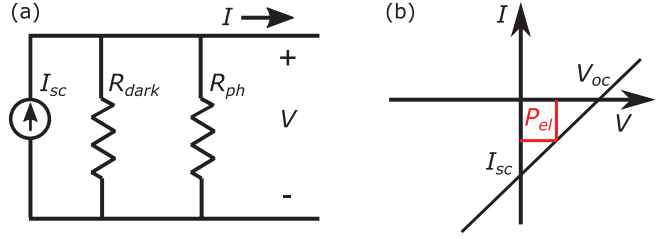}
    \caption{(a) Circuit diagram of a BPVE cell showing the current source $I_{sc}$ and parallel resistors $R_{dark}$ and $R_{ph}$ representing the dark conductivity, proportional to the intrinsic carrier density, and the photoconductivity, proportional to the photo-excited carrier density, respectively. (b) A resulting linear I-V curve with short circuit current $I_{sc}$, open circuit voltage $V_{oc}$ and the maximal extractable electrical power $P_{el}$, illustrating the fill factor of $25\%$.}
    \label{fig:circuitDiagram}
\end{figure}

In a good p-n junction solar cell, where the photovoltage represents a quasi-Fermi level separation in the absorber material, the I-V curve is the I-V curve of a diode, shifted by the light induced current.
In the BPVE, the photovoltage at open circuit is given by the balance of the shift current due to the light field and a drift current due to the static electric field generated by the voltage.
This can be represented by the simple circuit diagram in Figure \ref{fig:circuitDiagram}, which omits contact resistance \cite{Nakamura2018} or barriers forming at the contact, as we are interested in ideal limits.
The open circuit voltage, $V_{oc}$, is therefore inversely proportional to the conductivity $\sigma$ of the absorber material and also to the distance between contacts $d$, in stark contrast to the voltage in a p-n junction
\begin{equation}
    \label{eq:Voc}
    V_{oc} = \frac{J_{sc} d}{\sigma} = \frac{I_{sc} d}{\sigma A_{c}} \, , 
\end{equation}
where $J_{sc}$ is the short circuit current density.
The I-V curve is linear, as expected from the equivalent circuit diagram (Figure \ref{fig:circuitDiagram}).
The conductivity of the absorber material consists of an intrinsic component due to the carriers that are present in the dark, i.e. intrinsic carriers and carriers due to doping, and a photo-induced component due to the carriers generated through absorption of light.
Because a large open circuit voltage relies on a very small conductivity, only semiconductors with a small intrinsic carrier density are suitable candidates for energy conversion applications.

In most practical demonstrations in the visible and near UV regime, the photo-conductivity dominates over the intrinsic conductivity.
A comparison of the slopes of dark I-V and light I-V in the data shown in Figure 2 of \cite{Yang2010} shows this clearly.
Neglecting the intrinsic conductivity constitutes a zero-Temperature limit.
When considering solar energy conversion at room temperature, the intrinsic conductivity cannot generally be neglected.
As we show below, the intrinsic conductivity tends to dominate over the photo-conductivity in the infrared regime at room temperature.
It is only at absolute zero that the intrinsic conductivity may vanish for all bandgaps.

This intrinsic conductivity at finite temperatures introduces a temperature dependence into the efficiency limit.
Note that a temperature dependence is necessary for any efficiency limit to ensure adherence to the second law of thermodynamics.
In principle, radiative recombination also introduces a temperature dependence and its proper inclusion ensures that no currents occur in thermodynamic equilibrium \cite{Sturman2020} but we neglect this contribution as it is small at the bandgaps and device temperatures we consider.
In efficient BPVE devices, non-radiative recombination has to be much faster than radiative recombination for the photo-excited carrier density to remain small.

As illustrated in Figure \ref{fig:circuitDiagram}, the maximum power point of the linear I-V curve of the BPVE device is at half the open circuit voltage and half the short circuit current, resulting in
\begin{equation}
    \label{eq:powerLin}
    P_{BPVE} = V_{mpp} I_{mpp} = \frac{V_{oc} I_{sc}}{4} \, .
\end{equation}
which lets us write the efficiency at the maximum power point of the I-V curve as
\begin{equation}
    \label{eq:eff2}
    \eta_{tot} = \frac{V_{oc} I_{sc}}{4 \mathcal{I} A_{i}} \approx \frac{I_{sc}^2 d}{4 \mathcal{I} \sigma_{ph} A_{i} A_{c}} = \frac{G^{2} a^{2} \mathcal{I} A_{i}}{4 \sigma_{ph} d A_{c}}
    = \frac{q^{2} R^{2} a^{2} \mathcal{I} A_{i}}{4 d \hbar^{2} \omega^{2} \sigma_{ph} A_{c}} \, .  
\end{equation}

The photo-conductivity can be partially calculated from the band structure, a calculation to which we will return later, but here, we first make a few simple assumptions.
The first assumption is that the carriers in the conduction and valence band respectively are distributed according to a Fermi Dirac distribution at room temperature, which results from the reasonable assumption that carrier recombination is slower than carrier-phonon scattering.
Additionally, we make the parabolic approximation, which is usually a good approximation for carriers close to the band edge of a semiconductor.
The photo-conductivity $\sigma_{ph}$ can then be written as
\begin{equation}
    \label{eq:sigmaph}
    \sigma_{ph} = q (\mu_{e} n +\mu_{h} p) = q^{2} \tau_{sc} \left( \frac{n}{m_{e}} + \frac{p}{m_{h}}\right) = \frac{q^{2} \tau_{sc} \phi a \tau_{rec}}{m_{r} t} \, ,
\end{equation}
where $\tau_{sc}=(\tau_{rec}^{-1}+\tau_{c-ph}^{-1}+\tau_{c-defect}^{-1})^{-1}$ is the inelastic scattering lifetime, $\tau_{rec}$ is the recombination lifetime of the photo-generated carriers and t is the thickness of the material.
We assume here that the carrier mobility for electrons and holes respectively is given by $\mu_{e/h}=\tau_{sc} q/m_{e/h}$, so that the conductivity depends on the reduced mass of the semiconductor $m_{r}=(m_{e}^{-1}+m_{h}^{-1})^{-1}$, taken in the direction perpendicular to the contacts.

Inserting this result for the photoconductivity into the efficiency formula \eqref{eq:eff2} leaves us with
\begin{equation}
    \label{eq:effSimple}
    \eta_{tot} = \frac{m_{r} R^{2} a \mathcal{I} A_{i} t}{4 d \hbar^{2} \omega^{2} \tau_{sc} \tau_{rec} \phi A_{c}} = \frac{m_{r} R^{2} a}{4 \tau_{rec} \tau_{sc} \hbar \omega} \, ,
\end{equation}
where the absorptivity $a$ of the active material in the device is included, a quantity that depends on the details of the device structure as well as material properties.
We can also look at the efficiency per absorbed light power, which is arguably also a meaningful measure since non-absorbed light could still be converted by another device.
We then get an efficiency that is independent of the thickness of the absorber material
\begin{equation}
    \label{eq:effAbsorbed}
    \eta_{norm} = \frac{m_{r} R^{2}}{4 \tau_{rec} \tau_{sc} \hbar \omega} = \frac{m_{r} G^{2} \hbar \omega}{4 q^{2} \tau_{rec} \tau_{sc}} \, .
\end{equation}
Note that this corresponds to the formula arrived at in \cite{Zenkevich2014}, apart from the important factor of $4$ in the denominator which stems from the fill factor of a linear I-V curve.

Most importantly, the efficiency is proportional to the square of the average displacement $R$ of absorbed carriers.
For the shift current, this corresponds to the average shift vector while it corresponds to a vectorial average over the ballistic transport distance for the injection current.

\section{The zero-conductivity or ultimate limit}

There is no known bound on the magnitude of the average shift vector $R$ \cite{Tan2019}, so that equation \eqref{eq:effAbsorbed} may lead us to conclude that either the efficiency is unbounded, which would violate the second law of thermodynamics, or that there must be an as yet unspecified bound on the magnitude of the shift vector.
Consider, however, that, in calculating the shift vector from the bandstructure of the material, there is an implicit assumption that the shift vector - derived from electronic wavefunctions - is not influenced by the static electric field across the device.
Yet for large static electric fields, the shift vector must depend on the magnitude of the electric field across the device.
The maximum shift of charge $R_{max}$ that a photon of energy $\hbar \omega$ can affect in a material with bandgap $E_{g}$ against a static electric field is given by energy conservation as
\begin{equation}
    \label{eq:Rultimate}
    R_{max} = \frac{\hbar \omega - E_{g}}{q E_{static}} \, .
\end{equation}
Note that energy conservation places no restriction on the shift vector in the absence of a static electric field (at short circuit) as no work is performed by the shift current in this case.
There is therefore no contradiction with calculations showing finite shift currents at the band edge as, for example, in \cite{Cook2017}.

It follows from equation \eqref{eq:Rultimate} that there is an energy conservation limit on the efficiency that is independent of the conductivity of the material and it is given simply by
\begin{equation}                 \label{eq:ultimateMono}
 \eta_{\text{ultimate}} = \frac{\hbar \omega - E_{g}}{\hbar \omega} \, .
\end{equation}
In other words, the BPVE can only convert the above-bandgap portion of the absorbed photon energy into electrical work, performed by shifting the charge against the static electric field.
This is in stark contrast to the barrier layer photovoltaic effect, which can only convert the below bandgap portion of the absorbed photon energy into work.
Note also that the value of this limit may be larger than the fill factor of $25\%$ of the I-V curve.
This is in contrast to the efficiency of barrier layer photovoltaics which is always smaller than the fill factor (which may however be arbitrarily close to $1$) because the open circuit voltage is smaller than the absorbed photon energy and the photon flux limits the short circuit current.

This limit assumes that the shift vector/ballistic transport vector points in the same direction for each electron-hole pair generated by light absorption.
In the case of the injection current this would require a bandstructure in which all transport of photo-excited carriers is unidirectional.
Yet, the directional average of the group velocity over all states in the bandstructure at a particular energy $E$ vanishes because $E(\mathbf{k})=E(-\mathbf{k})$ must be fulfilled in a time-reversal symmetric material.
Therefore, such an ideal condition for large injection currents appears impossible and estimates of the asymmetry parameter range from $10^{-1}$ to $10^{-3}$~\cite{Sturman2020}.
Given the idealisations involved, it is unlikely that the ultimate limit can be approached.
In Section \ref{sec:stateOfTheArt} we discuss the state of the art of energy conversion efficiency and misconceptions that have arisen.
In the next Section we discuss how another, lower limit can be derived for a BPVE originating from the injection current, i.e., from the asymmetric carrier excitation with subsequent ballistic transport.

\section{Ballistic/injection current limit} \label{sec:ballistic}

Let us consider equation \eqref{eq:effAbsorbed} if the average displacement $R$ is due to ballistic transport of photo-excited carriers with an asymmetric momentum distribution.
For this we introduce a coefficient $\zeta$ that quantifies this asymmetry and is given by $\zeta=0$ for a perfectly symmetric initial momentum distribution and $\zeta=1$ for a unidirectional initial momentum distribution.

Using the parabolic band approximation, as we did for the derivation of the photo-conductivity, the group velocity of charge separation of the electron and hole excited by a photon of frequency $\omega$ is given by
\begin{equation}
    \label{eq:groupVel}
    v_{g} = \sqrt{\frac{2(\hbar \omega-E_{g})}{ m_{r}}} \, ,
\end{equation}
with the reduced mass $m_{r}$ in the direction perpendicular to the contacts.
The ballistic transport length is
\begin{equation}
    \label{eq:l0}
    l_{0} = v_{g} \tau_{sc} \, ,
\end{equation}
resulting in 
\begin{equation}
    \label{eq:Rfroml0}
    R(\omega) = l_{0}(\omega) \zeta(\omega) = \tau_{sc} \sqrt{\frac{2(\hbar \omega-E_{g})}{ m_{r}}} \zeta(\omega) \, .
\end{equation}
Therefore, long scattering times are beneficial to achieve large short circuit currents.
Large voltages, however, require a large resistance, and therefore short scattering times, a contradictory requirement. 

Inserting the expression for the transport length into equation \eqref{eq:effAbsorbed} gives a simple expression for the energy conversion efficiency of absorbed monochromatic light
\begin{equation}
    \label{eq:monoBallistic}
    \eta_{ballistic}(\omega) = \frac{1}{2} \zeta^{2} \frac{\tau_{sc}}{\tau_{rec}}\frac{\hbar \omega - E_{g}}{\hbar \omega} \, .
\end{equation}
Since $\tau_{sc} \leq \tau_{rec}$, this means that maximally half of the kinetic energy of the excited carriers can be converted to electric power in the ballistic effect.
Significantly, and as for the ultimate limit, none of the photon energy used to overcome the band gap of the material is converted to electrical power.
The monochromatic energy conversion efficiency of the BPVE is thus significantly lower than the monochromatic energy conversion efficiency for an optimised p-n junction, which can theoretically approach unity for high light intensity \cite{Green2001}.

Note that we can calculate the I-V curve for ideal asymmetry and in the absence of scattering other than carrier recombination ($\tau_{sc}=\tau_{rec}$) also for non-parabolic dispersions (the calculation procedure is described in Appendix A).
A pure third order dispersion of the form $E_{kin}(k) = a |k|^3$ would then result in a slightly higher monochromatic conversion efficiency of the kinetic energy of the carriers of $56.44\%$ compared to $50\%$ for a parabolic dispersion, while a Dirac dispersion ($E_{kin}(k)=a |k|$) would result in a lower value of $37.3\%$. In the following we consider only the parabolic dispersion.

\begin{figure}
    \centering
    \includegraphics{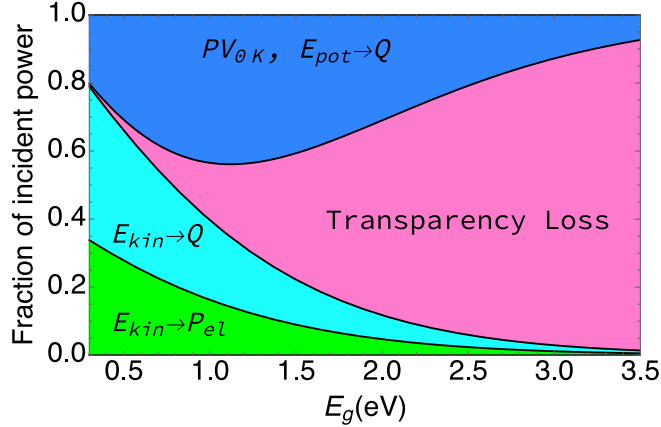}
    \caption{Zero temperature broadband limit of solar energy conversion with the injection current. The area labelled $PV_{0K}, E_{pot} \rightarrow Q$ describes the power that an ideal single junction photovoltaic device operated at $0$K could produce (in a mechanical analogy it could be seen as the potential energy of the carriers in the conduction band).
    The area labelled Transparency Loss describes the loss through photons that are not absorbed by the device.
    The area labelled $E_{kin} \rightarrow Q$ describes the part of the kinetic energy of the photo-excited carriers that cannot be converted to electrical power and is converted to heat, while the area labelled $E_{kin} \rightarrow P_{el}$ describes the electrical power that would be extracted by an ideal bulk photovoltaic with perfect excitation asymmetry.}
    \label{fig:zeroTempBroadbandLimit}
\end{figure}

For broadband excitation, the short circuit current is obtained through a frequency integral
\begin{equation}
    I_{sc}^{bb} = \int_{0}^{\infty} d\omega \frac{q \tau_{sc}}{d} \sqrt{\frac{2(\hbar \omega-E_{g})}{m_{r}}} A_{i} a(\omega) \phi(\omega) \zeta(\omega) \, .
\end{equation}
We describe the photon flux of unconcentrated sunlight through
\begin{equation}
    \label{eq:1SunPhotonFlux}
    \phi(\omega) = \frac{f}{h^{2} c^{2}} \frac{(\hbar \omega)^{2}}{e^{\frac{\hbar \omega }{k T_{Sun}}}-1} + \frac{ (1-f)}{h^{2} c^{2}} \frac{(\hbar \omega)^{2}}{e^{\frac{\hbar \omega }{k T_{env}}}-1} \, ,
\end{equation}
where $T_{Sun}=6000$K, the temperature of the environment is set to $T_{env}=300$K, and $f=1/46260$ describes the fraction of the sky taken up by the Sun. 
The open circuit voltage is again calculated from equation \eqref{eq:Voc}
where the conductivity $\sigma$ is given by
\begin{equation}
    \sigma = q^{2} \tau_{sc} n/m_{r} \, .
\end{equation}
The carrier density n is given by the sum of intrinsic carrier density $n_{i}$ and photo-generated carrier density
\begin{equation}
    n_{ph} = \tau_{rec} \int_{0}^{\infty} d\omega \phi(\omega) a(\omega) /t \, .
\end{equation}
Using equation \eqref{eq:powerLin} and dividing the result by the incoming energy flux density from the Sun
\begin{equation}
    \label{eq:1SunEnergyFlux}
    \dot{E} = \frac{f}{h^{2} c^{2}} \int_{0}^{\infty} d\omega \frac{(\hbar \omega)^{3}}{e^{\frac{\hbar \omega }{k T_{Sun}}}-1} \, ,
\end{equation}
we can derive a broadband zero-temperature limit by making the ideal but highly unrealistic assumption that $\tau_{sc}=\tau_{rec}$ and that $\zeta=1$, and by assuming a vanishing intrinsic carrier density.
The result is shown in Figure \ref{fig:zeroTempBroadbandLimit}
where we divide up the source of the different losses compared to an ideal $100\%$ conversion efficiency, inspired by the same plot for a single junction solar cell in \cite{Hirst2011}.
Interestingly, the BPVE uses none of the energy that is converted by the conventional single junction solar cell, as that is the below bandgap portion of the photon energy.
It uses some of the kinetic energy of the electrons instead, which is lost to thermalization in a single junction solar cell.
Since the maximum power voltage is different for different frequencies the maximum power voltage of a broadband device is a compromise between the different frequencies.
Therefore slightly less than half of the average kinetic energy of the photo-excited carriers can be converted to electrical energy, even if $\tau_{sc}=\tau_{rec}$, compared to half of the kinetic energy in the monochromatic case.

\section{Room temperature solar efficiency limit}

\begin{figure}
    \centering
    \includegraphics{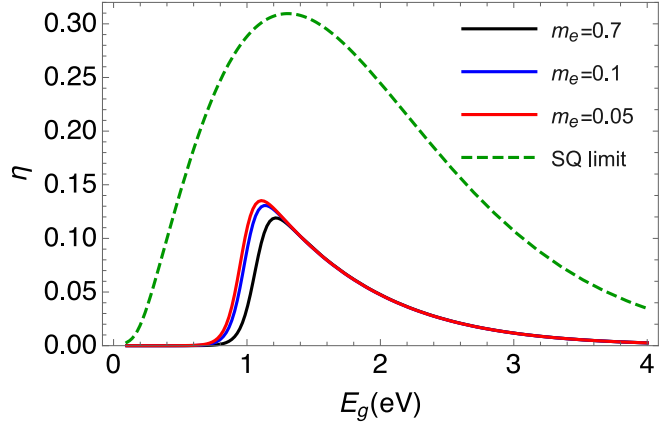}
    \caption{Room temperature efficiency limit for solar energy conversion assuming that the asymmetry of injected carriers is perfect ($\zeta=1$) and assuming different effective electron masses $m_{e}$ at the bandedge. The hole effective mass is set to $m_{h}=1$. The Shockley Queisser (SQ) limit is shown for comparison.}
    \label{fig:roomTempSolarLimit}
\end{figure}

At room temperature, we need to also consider the dark conductivity of the material.
To estimate its impact we can use some heuristic rules for the density of states of common semiconductors that give an order of magnitude estimate of the intrinsic carrier density at room temperature.
For materials with a bandgap in the blue region of the spectrum, the intrinsic carrier density is negligible compared to the photo-generated carrier density at solar intensities, therefore the dark conductivity is much smaller than the photo-conductivity.
This is, however, not necessarily true for materials with bandgaps in the infrared regime.
In Figure \ref{fig:roomTempSolarLimit} we now consider the dark conductivity at $300$K that arises from the intrinsic carrier density
\begin{equation}
\label{eq:intrinsicCarriers}
    n_{i} = \sqrt{N_c N_v} e^{\frac{-E_g}{2 k T}} \, ,
\end{equation}
that we calculate for different electron effective masses, assuming $m_{h}=1$, using the effective density of states $N_{c,v}$
\begin{equation}
    N_{c,v} = 2 \left( \frac{2 \pi m_{e,h} m_{0} k_B T }{h^{2}} \right)^{3/2} \, .
\end{equation}
The intrinsic carrier number depends on the thickness of the material, as does the absorptivity.
To connect absorptivity and intrinsic carrier density we use the approximate relation between bandgap and optical matrix element derived from the Thomas-Reiche-Kuhn sum rule \cite{YuCardonaBook}
\begin{equation}
\label{eq:OMETRK}
    p_{cv}^{2} = \frac{m_{0} E_{g}}{2} (m_{0}/m_{e}-1) \, .
\end{equation}
The absorption coefficient of the material is given by
\begin{equation}
    \label{eq:absorptionCoeff}
    \alpha(\omega) = \frac{\pi^{2} q^{2}}{n_{r} c \epsilon_{0}m_{0}^{2}\omega} \vert p_{cv} \vert^{2} N_{J}(\hbar \omega-E_{g}) \, ,
\end{equation}
with the joint density of states as function of kinetic energy $\varepsilon_{k}$
\begin{equation}
    N_{J}(\varepsilon_{k}) = \frac{1}{2 \pi^{2}} \bigg(\frac{2 m_{r}}{\hbar^{2}}\bigg)^{3/2} \sqrt{\varepsilon_{k}} \, .
\end{equation}
We choose the refractive index as $n_{r}=3.5$ and a thickness of the device of $t=2\mu$m.

The efficiency is independent of the intrinsic carrier density for large bandgaps but remains small overall because of the large transparency losses, as well as the small kinetic energy of the carriers.
The addition of the dark conductivity starts to suppress the achievable efficiency at a bandgap of around $1$eV.
This is due to a collapse of the photo-voltage because of the decreased resistance across the device.
The exact photon energy at which the collapse occurs depends on the value of the effective mass that we choose and would also change slightly if we chose a different refractive index.
However, the dependence on both of these factors is not strong because the dependence of the dark conductivity on the bandgap is exponential compared to the geometrical dependence on the reduced effective mass and refractive index (see equation \eqref{eq:intrinsicCarriers}).
We can clearly see that, even for the most optimistic and unrealistic assumptions, the energy conversion efficiency of the BPVE is much smaller than the Shockley Queisser limit.

\section{Efficiencies of existing materials} \label{sec:stateOfTheArt}

Our results for the energy conversion efficiency limits of the BPVE contradict several claims of high energy conversion efficiency that have been made in the last decade.
In the following we explain for each of the claims why we believe that they are erroneous. 

Let us start by noting that it was a long established consensus that the BPVE is an inefficient means of energy conversion.
Equation \eqref{eq:effAbsorbed} for the absorbed light contains a factor of ballistic transport (scattering) time over recombination time.
In \cite{Fridkin2001}, Fridkin gives an estimate for the typical ratio of $\tau_{sc}/\tau_{rec} \approx 10^{-4}-10^{-6}$,
limiting the BPVE to even smaller values as can be seen by inserting this ratio into equation \eqref{eq:monoBallistic} for the ballistic efficiency limit.

However, a later experimental claim of an efficiency of $0.6\%$ \cite{Zenkevich2014} was made by Fridkin and co-authors.
In that work, a $20$nm and a $50$nm BaTiO3 film are illuminated with monochromatic light at a frequency of $3.4$eV and the current flowing through Pt electrodes located above and below the film is measured.
The product of measured short circuit current densities $J_{sc}$ and open circuit voltage $V_{oc}$ is then normalised to the estimated absorbed light intensity to obtain an efficiency.
Thereby, the authors omitted the fill-factor of the J-V curve, the inclusion of which decreases the actual efficiency by a factor of $4$.
The authors also seem to have underestimated the absorption coefficient of the material, given in their work as $\alpha=5 \times 10^{2}/$cm, by a factor of at least $3$ but possibly by a much larger factor, as can be seen by comparison to \cite{Casella1959,DiDomenico1968}.
Also, because of the nanoscale thickness of the device which introduces optical resonances, it is not appropriate to calculate the absorption in the active material by simply multiplying the absorption coefficient with the thickness, an approach only valid for bulk materials.
Taking all this into account it is plausible that the measured efficiency is actually below $10^{-5}$, consistent with the earlier predictions by Fridkin \cite{Fridkin2001}.

\cite{Spanier2016} used a nanoscale tip to also measure the BPVE in BaTiO3 and claimed an extraordinary efficiency under AM1.5G illumination of $4.8\%$, which surpasses the Shockley Queisser limit for the same ($3.2$eV)  bandgap. 
The authors of \cite{Spanier2016} divided the power obtained from the I-V curve of their nanoscale tip not by the total light power impinging on the device, as is the standard definition of power conversion efficiency, but rather assumed a ballistic collection area of the tip and calculated the relevant light power using this area.
This implicitly assumes that a tightly focused spot of light around the nanotip would have produced the same current and voltage, which has not been shown by the authors.
Therefore \cite{Kirk2017} rightly pointed out that the normalisation procedure is inadmissible and the actual power conversion efficiency of the device investigated in \cite{Spanier2016} was $1.2 \times 10^{-8}$.

\begin{figure}
    \centering
    \includegraphics{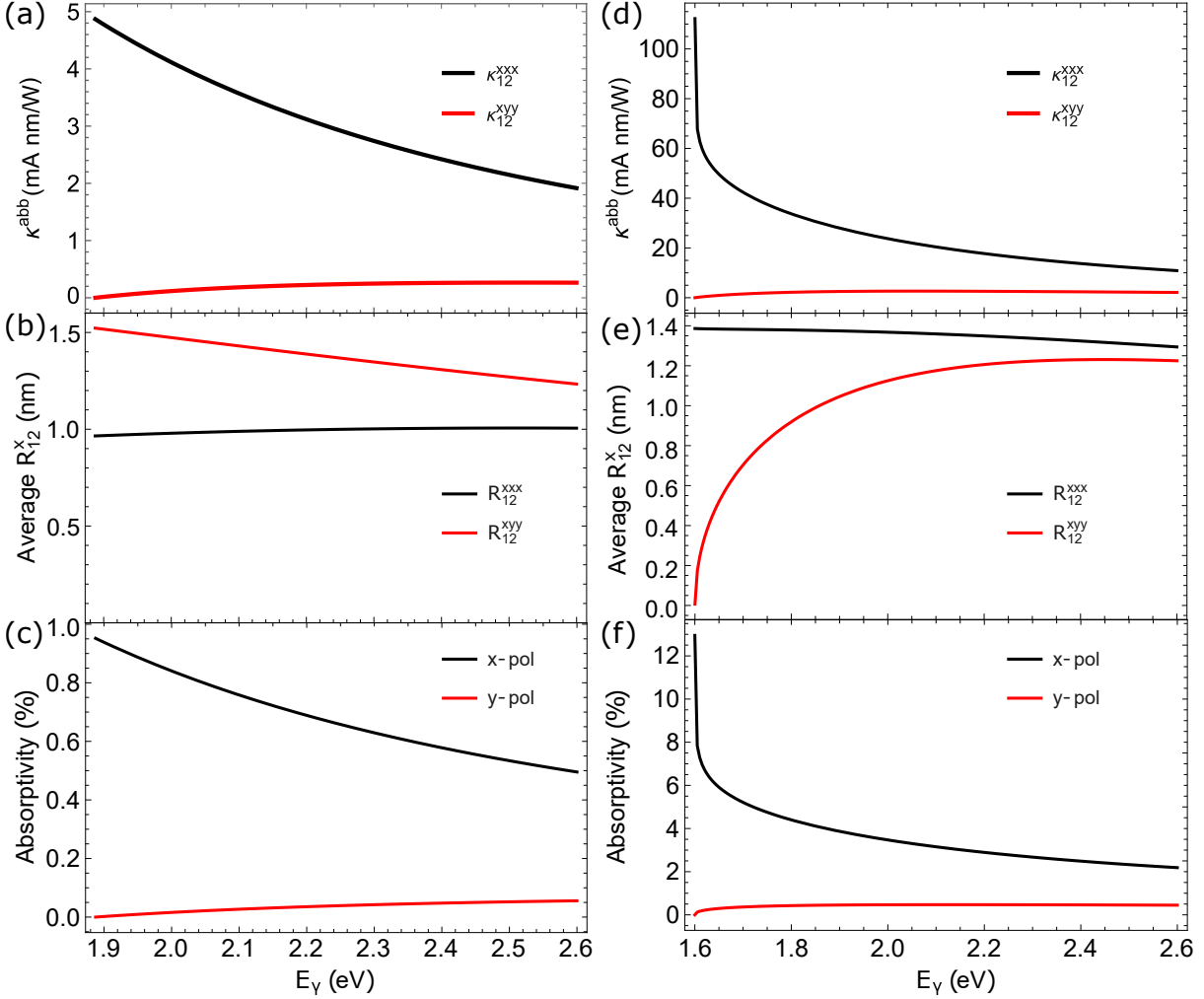}
    \caption{Frequeny dependent (a) photoresponsivity, (b) average shift vector and (c) absorptivity of 2D GeSe. Frequency dependent (d) photoresponsivity, (e) average shift vector and (c) absorptivity of a 2D semi-Dirac material.}
    \label{fig:2Dplots}
\end{figure}

Recently, various 2D materials have been put forward as potentials materials for energy conversion using the shift current mechanism \cite{Cook2017}.
Advocates for 2D materials point out that the 2D photoresponsivity can be converted to a remarkably high 3D value if the responsivity is divided by the thickness of the material.
However as we established with equation \eqref{eq:effAbsorbed}, the relevant parameter for high conversion efficiency is the average shift vector.
The average shift vector for a particular frequency and polarisation is the average shift upon absorption of a photon and can be calculated by dividing the total electrical current $J_{a} w$ by the total rate of absorbed photons and the unit charge
\begin{equation}
    R_{abb} = \frac{J_{a} w}{a q \phi_{b} w d q} = \frac{\kappa_{abb} \hbar \omega} {q a} \, .
\end{equation}
where $\kappa_{abb}$ is the photoresponsivity (for details of the calculation see the Appendix B) of the material leading to a current density (in units of $A/m$ because the current flows in a 2D material) in direction $a$ from a light polarised in direction $b$, i.e, $J_{a}=\kappa_{abb} \mathcal{I}(b)$ and $a$ is the absorptivity of the material for this particular polarisation.
Figures \ref{fig:2Dplots}(a) and (d) show those photoresponsivities for GeSe and a semi-Dirac material with otherwise similar properties that has been designed for maximum photoresponsivity (see Appendix for details on the model Hamiltonian).
We show the 2D photoresponsivity, which retains an additional unit of nm compared to the photoresponsivity for a 3D material, because the current density is a density across a line of material, not along an area.

As can be seen in Figure \ref{fig:2Dplots}(b) and (e), the average shift vector in GeSe and related materials is within the usual range for bulk materials, on the order of a nanometer.
There are a few interesting observations to be made here.
Firstly, while the susceptibility is much larger for $x$-polarized than for $y$-polarized light, this is exclusively due to the differences in absorptivity (see Figure \ref{fig:2Dplots}(c) and (f)) for the two linear polarizations.
In fact, the shift vector is larger for $y$-polarized light despite the susceptibility being much larger for $x$-polarized light.
Therefore, the total power obtained from a single layer of GeSe is larger when it is illuminated by $x$-polarized light but the conversion efficiency of absorbed light is larger for $y$-polarized light.
Therefore, the total power obtained from a large number of 2D GeSe layers may be higher for $y$-polarized light.
This underlines that the second order susceptibility itself is not a reliable indicator of energy conversion efficiency.

To convert the shift vector into an energy conversion efficiency $\eta_{abs}$ using equation \eqref{eq:effAbsorbed} we need to make assumptions about the scattering and recombination lifetimes in the semiconductor.
The smaller the recombination lifetime and the smaller the scattering lifetime, the larger the efficiency returned by \eqref{eq:effAbsorbed}.
Using optimistically low values for the recombination lifetime of $1$ps and the scattering lifetime of $10$fs, a photon energy of $2$eV, and the effective mass in $x$-direction of $0.643 m_0$ resulting from the tight binding Hamiltonian in \cite{Cook2017} we arrive at a monochromatic conversion efficiency for absorbed light of $\eta_{abs}^{GeSe} =4.5 \times 10^{-5}$ for a shift vector of $~1$nm in 2D GeSe.
The semi-Dirac material shows a much higher photoresponsivity which, however, does not correspond to a much higher shift vector at a peak of $1.4$nm. The efficiency $\eta_{abs}^{SD}=3.05 \times 10^{-5}$ calculated from equation \eqref{eq:effAbsorbed} is actually lower than for GeSe because its effective mass in $x$-direction is smaller at $m_{eff,x}^{SD}=0.219 m_{0}$ (see Appendix).

Those efficiencies are four orders of magnitude too low to compete with p-n junction photovoltaics on energy conversion applications.

Finally, another recent work found a very high Glass coefficient for TaAs \cite{Osterhoudt2019}, which has a bandgap in the mid-infrared region.
Noting that the energy conversion efficiency for a given Glass coefficient is proportional to the excitation frequency at which this Glass coefficient is obtained, it is clear that this may not necessarily translate into higher energy conversion efficiency.
More importantly, as we have shown in Figure \ref{fig:roomTempSolarLimit}, the intrinsic conductivity cannot be neglected in the mid-infrared regime, and large voltages may be impossible to obtain.

Having clarified that the operating efficiencies for BPVE devices to date are low, we can still use equation \eqref{eq:effAbsorbed} to consider what needs to be achieved to obtain higher energy conversion efficiencies.
For the shift current we need a material with a high average shift vector that has a very low conductivity.

The injection current relies on a high asymmetry factor $\zeta$.
Fast scattering times increase the resistance and thereby the voltage but also decrease the ballistic travel distance and therefore the short circuit current, negating the advantage.
Ultimately, however, it seems that energy conversion is not the right application for this fascinating effect.
As mentioned above, this was already recognised a few decades ago \cite{Fridkin2001} and recent experimental and theoretical results have not changed this picture.

\section{Conclusion}

Photovoltages that exceed the bandgap energy in non-centrosymmetric materials, arising from second order nonlinearities have raised hopes for a novel means to surpass the Shockley Queisser limit for solar energy conversion.
We show that, while the BPVE is not subject to the Shockley Queisser limit, its conversion mechanism has more stringent limitations, even with the most generous assumptions.
In particular, by employing the most generous bounds possible, we showed that the solar energy conversion efficiency for the BPVE is bounded by slightly less than half of the kinetic energy portion of the absorbed photons, i.e. the photon energy that is larger than the energy needed to excite electrons from valence to conduction band.
In that sense, the BPVE constitutes a counterpart to the barrier layer device in that it uses only the part of the photon energy that the barrier layer device cannot use.
Note that this does not mean that the two effects can be easily combined in the same device as the BPVE requires very low conductivity and high nonradiative recombination rates to achieve appreciable voltages whereas the barrier layer device requires high conductivity and low nonradiative recombination rates to achieve good performance.

The large photovoltage of the BPVE relies on the insulating properties of the absorbing medium experienced with large bandgap materials.
In the zero-temperature limit, where intrinsic carriers cease to exist, this could lead to efficiencies beyond the Shockley Queisser limit, however, useful solar energy conversion occurs at temperatures around $300$K, where the conductivity of any low bandgap material becomes high and the photovoltage collapses.
We showed that this collapse occurs near a bandgap of $1$eV for solar light intensities and this cutoff energy only weakly depends on the assumed effective mass near the bandgap.

We explained that confusion in the recent literature regarding high energy conversion efficiencies is mainly due to incorrect normalisation procedures for absorption.
In one prominent case \cite{Spanier2016} the light to electricity energy conversion efficiency is calculated by normalising the electrical output power by only a small fraction of the total incoming energy flux in the radiation field, something that was already been pointed out in \cite{Kirk2017} but has largely been overlooked by the BPVE community.
Another experimental suggestion of large efficiency \cite{Zenkevich2014} relied on severely underestimating the absorption coefficient of the underlying material and thus normalising the electrical output by the incorrect quantity of absorbed radiation.
Finally, the rise of 2D materials has lead to a large body of literature on giant susceptibilities in such monolayers.
These susceptibilities appear giant because they are compared to 3D material susceptibilities by normalising them by dividing by the (extremely small) thickness of the monolayer instead of normalising by the (appreciable) absorptivity of such a layer.
If $~1\%$ of the incoming light is absorbed by the monolayer, the shift current of an effective 3D material made of a large number $N>100$ of such monolayers, is not equal to $N$ times the susceptibility of the monolayer, simply because of light attenuation.

We use the particular example of a GeSe monolayer to illustrate that the efficiencies for 2D materials is not expected to be much higher than the efficiencies of well-known 3D materials.
This leads us to conclude that the efficiency estimate made by \cite{Fridkin2001} still holds; the energy conversion efficiency of the BPVE is given by approximately $10^{-5}$ to $10^{-6}$ for good materials.
None of the recently proposed materials are significantly more efficient than this and there is no reason to expect that BPVE devices with energy conversion efficiencies comparable to those of commercial p-n junction solar cells are possible.

\section*{Appendix}

\subsection{Calculation procedure for the ideal ballistic conversion efficiency}

To calculate the ideal ballistic conversion efficiency of kinetic energy into electrical energy for an arbitrary dispersion, mentioned in Section \ref{sec:ballistic}, we need to calculate the average distance $R$ travelled by an electron with a given initial kinetic energy $K_{0}$ against an electric field $E$.
Assuming random recombination we thus have
\begin{equation}
\label{eq:ballisticAppendix}
    R = \int_{0}^{\infty} dt e^{-t/\tau_{rec}} v_{g}(t) \, .
\end{equation}
$v_{g}(t)$ for ballistic transport is obtained by first calculating $v_{g}(K(t))$ from the dispersion relation.
The kinetic energy as function of time is given by
\begin{equation}
    K(t) = K_{0} - q E x(t) \,
\end{equation}
for the forward part of the ballistic trajectory and
\begin{equation}
    K(t) = q E \big(x_{0}-x(t)\big) \,
\end{equation}
for the backward part of the trajectory.
Here, $x_{0} = K_{0}/(q E)$ is the point at which the electron has lost all of its initial kinetic energy.

Transforming the equation for $K(t)$ into an equation for $v_{g}(t)$ we obtain a differential equation for $x(t)$.
The I-V curve is traced by varying the electric field (which is proportional to the voltage) and calculating $R$ (which is proportional to the current) from equation \eqref{eq:ballisticAppendix} inserting $v_{g}(K(x(t)))$.

\subsection{Second order susceptibility and absorptivity of a 2D material} \label{sec:SecondOrderSuscepDetails}

To evaluate the potential for 2D GeSe and the 2D semi-Dirac material we use the low-energy approximation to the tight binding Hamiltonian $H$ presented in \cite{Cook2017}
\begin{equation}
    H = \sigma_{0} \epsilon_{0} + \sum_{i} \sigma_{i} f_{i} \, ,
\end{equation}
where $\sigma_{0}$ is the identity matrix and the $\sigma_{i}$ are the Pauli matrices.
$\epsilon_{0}$ and the $f_i$ are wave-vector dependent parameters of the Hamiltonian.

The shift current density (in units of $A/m$ as the current flows in a 2D material) in direction $a$ $J_a$ for linear polarized light is calculated from the $2^{nd}$ order susceptibility $\sigma^{abb}$ as
\begin{equation}
    J_{a} = \sigma^{abb} E_{b}(\omega) E_{b}(-\omega) \, ,
\end{equation}
where $E_{b}$ is the electric field of the light polarized along direction $b$.
Alternatively, this can be written in terms of the photoresponsivity $\kappa^{abb}$, where $J_{a}=\kappa^{abb} \mathcal{I}_{b}$ with the light intensity $\mathcal{I}_{b}$.
According to \cite{Cook2017}, the photoresponsivity for a 2D system is
\begin{equation}
    \kappa^{abb} = \frac{4 g_{s} \pi q^{3}}{\hbar c \epsilon_{0}} \int \frac{d^{2}k}{(2 \pi)^{2}} f_{nm} I^{abb}_{nm} \delta(\hbar \omega_{nm}- \hbar \omega),
\end{equation}
where $\omega_{nm}$ is the wave-vector dependent resonance frequency between band $n$ and $m$, while $g_{s}=2$ is the spin degeneracy and $f_{nm}$ is the difference between occupation densities of the electronic states.
We only consider non-degenerate semiconductors, where $f_{12}\approx 1$.

For the Hamiltonian under consideration, the integrand determining the shift current can be written as \cite{Cook2017}
\begin{equation}
    I^{abb}_{12} = - \sum_{ijm} \frac{1}{4 \epsilon^{3}} \left( f_{m}f_{i,b}f_{j,ab} - f_{m} f_{i,b} f_{j,a} \frac{\epsilon_{,b}}{\epsilon} \right) \varepsilon_{ijm} \, .
\end{equation} 
Here, the subscript $,a$ denotes the derivative towards $k_{a}$, $\epsilon = \sqrt{\sum f_{i}f_{i}}$, and $\varepsilon_{ijm}$ is the Levi Civita symbol.

For an electron-hole symmetric material the absorptivity for light polarized along direction $a$ is calculated as
\begin{equation}
\label{eq:absorptivities2D}
\frac{q^{2} \omega}{ 2 \pi \hbar c \epsilon_{0}} \int d^{2}k 
  \vert r^{a}_{21}(\mathbf{k}) \vert^{2} \delta\left(\hbar \omega_{12}(\mathbf{k})-\hbar \omega\right) \, ,
\end{equation}
with the $a$-component of the dipole matrix element $r^{a}_{21}$, the resonance frequency $\omega_{12}(\mathbf{k})$ and the Dirac delta function $\delta$.
For the Dirac Hamiltonian of graphene this results in the famous frequency-independent absorptivity of $2.3\%$.

\subsection{Parameters for 2D GeSe and semic-Dirac materials} \label{sec:parameters2D}

The momentum dependent functions $f_{i}$ are given by
\begin{equation}
    f_{x} = \delta + \alpha_{x} k_{x}^{2} + \alpha_{y} k_{y}^{2} + \alpha_{xy} k_{x} k_{y} \, ,
\end{equation}
\begin{equation}
    f_{y} = v_{f} k_{x} \, ,
\end{equation}
and
\begin{equation}
    f_{z} = \Delta + \beta_{x} k_{x}^{2} + \beta_{y} k_{y}^{2} + \beta_{xy} k_{x} k_{y} \, .
\end{equation}
This Hamiltonian is time-reversal symmetric but has broken inversion symmetry.

The tight binding Hamiltonian for a 2D layer of GeS is parameterized through
\begin{equation}
    f_{x} + i f_{y} = - e^{- i \mathbf{x}_{0} \mathbf{k}}\big[ t_{1} + t_{2} \Phi(k) + t_{3} \Phi^{\ast}(k) \big] \, ,
\end{equation}
with
\begin{equation}
    \Phi(k) = \big(e^{i \mathbf{a}_{1} \mathbf{k}} + e^{i \mathbf{a}_{2} \mathbf{k}}\big) \, ,
\end{equation}
and
\begin{equation}
    f_{z} = \Delta \, .
\end{equation}
$\mathbf{a}_{1}$ and $\mathbf{a}_{2}$ are the lattice vectors of the two-dimensional lattice, while $x_{0}$ is the distance between the  orbital centers of the two different sites of the unit cell.
This results in
\begin{equation}
    v_{f} = 2 a_{x} (t_2 - t_3) + (t_1 + 2 t_2 + 2 t_3) x_0 \, ,
\end{equation}
\begin{equation}
    \delta = -t_1 - 2 t_2 -  2 t_3 \, ,
\end{equation}
\begin{equation}
   \alpha_{x} = t_2 (a_x + x_0)^2 + t_1 x_{0}^{2}/2 + t_3 (a_x - x_0)^2 \, , 
\end{equation}
and
\begin{equation}
 \alpha_{y} = (t_2 + t_3) a_{y}^{2} \, ,
\end{equation}
while $\alpha_{xy} = \beta_{x} = \beta_{y} = \beta_{xy}=0$

The tight binding fit to the GeS bandstructure is given by \cite{Cook2017} as $\Delta = 0.41$eV, $t_{1}=-2.33$eV, $t_{2}=0.61$eV, $t_{3}=0.13$eV, $\mathbf{a}_{1}=(2.765,0)$\AA, $\mathbf{a}_{2}=(0,1.815)$\AA, and $x_{0}=0.62$\AA.
In order to calculate the efficiency of energy conversion from absorbed light to electricity we also need the effective masses in each direction, given by 
\begin{equation}
    m_{eff,x} = \frac{2 q^{2}}{\hbar^{2}} \frac{v_{f}^{2} + 2 \alpha_{x} \delta + 2 \beta_{x} \Delta }{E_{g}} = 0.643 m_{0} \, ,
\end{equation}
and
\begin{equation}
    m_{eff,y} = \frac{4 q^{2}}{\hbar^{2}} \frac{\alpha_{y} \delta + \beta_{y} \Delta}{E_{g}} = 1.735 m_{0} \, ,
\end{equation}
for this model and the parameters for GeSe.

To model a semi-Dirac material we replace the following parameters, $t_{2}=1.164$, $t_{3}=0$, $\Delta = 0.8$, and $x_{0}=0.6$\AA.
This results in effective masses of $m_{eff,x}=0.219 m_{0}$ and, as intended, an almost infinite mass in $y$-direction of $m_{eff,y}=397.5 m_{0}$.

\bibliography{main}
\end{document}